\documentclass[pdflatex,sn-aps,iicol]{sn-jnl}

\usepackage{graphicx}
\usepackage{xcolor}
\usepackage{amsmath,amssymb,mathtools}
\usepackage[title]{appendix}
\usepackage{booktabs}
\usepackage{cuted}
\newcounter{mainfigurecounter}

\hypersetup{colorlinks=true, linkcolor=blue, citecolor=blue, urlcolor=blue}

\allowdisplaybreaks[4]

\begin{document}

\articletype{Research Article}

\title{Probability-based Estimates of the Uncalculated N\texorpdfstring{$^5$}{5}LO QCD Contribution to the Hadronic \texorpdfstring{$W$}{W}-Boson Decay Width}
		
\author[1]{\fnm{Shu-Heng} \sur{Yang}}\email{yangsh@stu.cqu.edu.cn}

\author*[2]{\fnm{Jiang} \sur{Yan}}\email{yjiang@itp.ac.cn}

\author[1]{\fnm{Xing-Gang} \sur{Wu}}\email{wuxg@cqu.edu.cn}

\author[1]{\fnm{Zhi-Fei} \sur{Wu}}\email{wuzf@cqu.edu.cn}

\affil[1]{\orgdiv{Department of Physics, Chongqing Key Laboratory for Strongly Coupled Physics}, \orgname{Chongqing University}, \orgaddress{\city{Chongqing}, \postcode{401331}, \country{China}}}
\affil*[2]{\orgname{Institute of Theoretical Physics, Chinese Academy of Sciences}, \orgaddress{\city{Beijing}, \postcode{100049}, \country{China}}}

\abstract{
The perturbative QCD corrections to the hadronic decay width of the $W$ boson are currently known up to next-to-next-to-next-to-next-to-leading order ($\mathrm{N^4LO}$), whereas the exact $\mathrm{N^5LO}$ correction remains unavailable owing to its formidable computational complexity. In this work, we estimate the $\mathrm{N^5LO}$ contributions by adopting Bayesian analysis (BA). Before performing the estimation, the Principle of Maximum Conformality (PMC) is employed to improve the precision of the initial scale-dependent perturbative series. Through recursive application of the renormalization group equation, non-conformal terms are absorbed into the strong running coupling, yielding a scheme-independent, scale-invariant perturbative series with improved convergence. The PMC procedure determines an effective coupling $\alpha_s(Q_*)$, with the PMC scale fixed as $Q_* = 100.102~\mathrm{GeV}$ at next-to-next-to-leading logarithmic accuracy. Based on the improved and more precise series, the $95.5\%$ BA credible interval yields an uncertainty of $\Delta\delta_\mathrm{QCD}|_{\mathrm{PMC, BA}}^{\mathrm{N^5LO}} = \pm 1.0\times 10^{-5}$. The resulting hadronic branching ratio is $\mathcal{B}(W\to \text{hadrons})|_\mathrm{PMC} = (65.84\pm 1.54)\%$, which is consistent with experimental data within reasonable errors.	
}


\maketitle
		
\section{Introduction}
		
     The $W$ boson serves as a key probe of electroweak symmetry breaking and a critical benchmark for validating the internal consistency of the Standard Model (SM). Within the SM, the hadronic decay width of the $W$ boson depends on several fundamental input parameters, including the strong coupling constant $\alpha_s(M_W)$ evaluated at the $W$-boson mass scale and the quark-flavor mixing entries in the first two rows of the Cabibbo–Kobayashi–Maskawa (CKM) matrix. Precise theoretical predictions for the hadronic $W$-boson decay width are essential for stringent SM tests and for constraining potential contributions from physics beyond the SM. As collider measurements at facilities such as the LHC achieve ever-increasing precision, theoretical uncertainties must be systematically reduced to a comparable level and reliably quantified.
		
     The hadronic decays of the $W$ boson feature a large number of final states, rendering a direct inclusive measurement of the hadronic decay width experimentally challenging. The hadronic decay width is therefore typically extracted indirectly from the total and leptonic decay widths. In 2010, the combined Tevatron result from the CDF and D0 Collaborations yielded $\Gamma_W = 2046 \pm 49~\mathrm{MeV}$, based on the $W\to e\nu$ and $W\to \mu\nu$ decay channels~\cite{TevatronElectroweakWorkingGroup:2010mao}. In 2013, the ALEPH, DELPHI, L3, and OPAL Collaborations performed a combined analysis of LEP data collected over center-of-mass energies spanning $130$--$209~\mathrm{GeV}$ with an integrated luminosity of approximately $3~\mathrm{fb}^{-1}$, obtaining $\Gamma_W = 2195 \pm 83~\mathrm{MeV}$ and a hadronic branching fraction of $(67.41 \pm 0.27)\%$~\cite{ALEPH:2013dgf}. In 2022, the CMS Collaboration reported a measured hadronic branching fraction of $(67.32 \pm 0.23)\%$~\cite{CMS:2022mhs}. In 2024, the ATLAS Collaboration released the first LHC measurement of the $W$-boson decay width, giving $\Gamma_W = 2202 \pm 32\ (\mathrm{stat.}) \pm 34\ (\mathrm{syst.})~\mathrm{MeV}$~\cite{ATLAS:2024erm}. The current world-average value compiled by the Particle Data Group (PDG) is $\Gamma_W = 2140 \pm 50~\mathrm{MeV}$~\cite{ParticleDataGroup:2026aaa}.
		
    The hadronic $W$-boson decay width receives perturbative corrections from QCD, electroweak (EW), and mixed QCD--EW contributions, and can be expressed as
    \begin{align}\label{eq:Gamma}
    \Gamma_{W}^\mathrm{had} =& \frac{G_F N_c M_W^3}{6\pi\sqrt{2}} \sum_{q=u,c}\sum_{q'=d,s,b} |V_{qq'}|^2 \notag\\
    & \times \bigg[ 1 + \delta_\text{QCD}(\alpha_s) + \delta_\mathrm{EW}(\alpha) + \delta_\mathrm{mix}(\alpha,\alpha_s) \bigg],
    \end{align}
    where $N_c=3$ denotes the number of colors, $G_F=1.166379\times10^{-5}~\mathrm{GeV}^{-2}$ is the Fermi constant, $M_W$ is the $W$-boson mass, and $V_{qq'}$ represents the CKM matrix elements. The double summation runs over the quark flavors $q=u,c$ and $q'=d,s,b$. 
    
    Substantial progress has been made in calculating radiative corrections to the hadronic $W$-boson decay width. In the massless-quark limit, the QCD correction is known up to four-loop level ($\mathcal{O}(\alpha_s^4)$)~\cite{Gorishnii:1990vf, Surguladze:1990tg, Chetyrkin:1996ez, Baikov:2008jh}. For finite quark masses, the one-loop QCD corrections have been computed in Refs.~\cite{Chang:1981qq, Alvarez:1987gi}, while two- and three-loop QCD corrections including quark-mass contributions are available in Ref.~\cite{Chetyrkin:1996hm}. Furthermore, one-loop QED corrections for massless fermions were derived in Refs.~\cite{Marciano:1973rig, Albert:1979ix}; full one-loop electroweak (EW) corrections are known through $\mathcal{O}(\alpha)$~\cite{Inoue:1980ky, Consoli:1983yn, Bardin:1986fi}, and mixed QCD--EW corrections for finite fermion masses have been evaluated at $\mathcal{O}(\alpha\alpha_s)$~\cite{Denner:1990cpz, Denner:1991kt, Kniehl:2000rb, Kara:2013dua}.
    
    Numerically, relative to the Born-level width, the pure EW and mixed QCD--EW corrections contribute approximately $\delta_\mathrm{EW} \sim -0.36\%$ and $\delta_\mathrm{mix} \sim -0.05\%$~\cite{dEnterria:2016rbf}, respectively, while the finite-mass effect amounts to about $-0.12\%$~\cite{Chetyrkin:1996hm}. The QCD correction is therefore the dominant radiative contribution to hadronic $W$-boson decay and requires a precise theoretical treatment.
    
    The typical momentum flow governing hadronic $W$-boson decay is of order $M_W$. Given $\alpha_s(M_W) \sim 0.12$, higher-order terms are effectively suppressed by powers of $\alpha_s$, and the $\mathrm{N^4LO}$ pQCD series exhibits good convergence with reduced net scale dependence. Nevertheless, the initial fixed-order perturbative series still suffers from renormalization scale dependence arising from the mismatch between the magnitude of $\alpha_s$ and its expansion coefficients, while its convergence is degraded by divergent renormalon contributions~\cite{Beneke:1994qe, Neubert:1994vb}. In conventional analyses, the renormalization scale is usually set to $\mu_\mathrm{R} = M_W$ to remove large logarithmic corrections, and the scale uncertainty is simply estimated by varying $\mu_\mathrm{R} \in [M_W/\xi,\,\xi M_W]$ with $\xi = 2,3,\dots$. This ad hoc prescription introduces a systematic ambiguity through both the choice of central scale and the variation range. Moreover, the scale dependence of individual perturbative contributions complicates reliable estimates of unknown higher-order (UHO) terms, thereby limiting the precision of theoretical predictions.
	
    In principle, theoretical prediction for any physical observable should be independent of the renormalization scale~\cite{Callan:1970yg, Symanzik:1970rt, Peterman:1978tb}. Nevertheless, constrained by standard renormalization-group invariance, exact scale independence cannot be naively achieved for fixed-order series due to unknown higher-order (UHO) terms, necessitating a systematic scale-setting procedure. For non-Abelian QCD, the Principle of Maximum Conformality (PMC)~\cite{Brodsky:2011ig, Brodsky:2012rj, Brodsky:2011ta, Mojaza:2012mf, Brodsky:2013vpa} provides a rigorous framework for constructing fixed-order pQCD series consistent with renormalization-group principles~\cite{Brodsky:2012ms, Wu:2013ei, Wu:2014iba, Wu:2019mky}. Since the running of the QCD coupling is governed by the renormalization group equation (RGE), all RGE-relevant $\{\beta_i\}$ terms in the pQCD series can be resummed to build an effective coupling $\alpha_s$ within the PMC single scale-setting (PMCs) formalism~\cite{Shen:2017pdu, Yan:2022foz}. It is then verified that the PMC effectively eliminates the scheme and scale dependence of the initial pQCD series, yielding high-precision fixed-order predictions~\cite{Wu:2018cmb}. Accordingly, we adopt the PMCs approach throughout the present analysis. 
		
    Furthermore, since the complete all-order pQCD result remains unknown, reliable methods are required to quantify UHO contributions and their associated uncertainties. To this end, several probabilistic approaches have been proposed in the literature, such as the Bayesian analysis~(BA)~\cite{Cacciari:2011ze, Bagnaschi:2014wea, Bonvini:2020xeo, Duhr:2021mfd, Shen:2022nyr} and the linear regression through the origin~(LRTO)~\cite{Wu:2025xdv}. In the present work, we employ both approaches for our discussion, and we demonstrate below that the PMC conformal series yields a more stable coefficient sequence for probability-driven evaluations of UHO contributions. Bayesian analysis constitutes a powerful framework for constructing probability distributions, where the conditional probability of the UHO terms is first specified via a subjective prior distribution and then iteratively updated by means of Bayes' theorem upon incorporating additional information. In contrast to the flexible prior setup adopted in BA, the LRTO approach assumes a definite linear behavior for the logarithmic form of the series. A detailed order-by-order comparison shows that the LRTO estimate is unstable for the present observable, indicating that the present perturbative series does not support such linear scaling. We therefore adopt the BA result as our nominal estimate for the uncertainty originating from UHO terms in the current $\mathrm{N^4LO}$ prediction, while quoting the LRTO result solely for comparison.

    The remainder of this paper is organized as follows. \autoref{sec:th-fw} introduces the theoretical framework, including the $\mathrm{N^4LO}$ pQCD correction to the hadronic $W$-boson decay width, the PMC scale-setting procedure, and the BA method adopted to estimate UHO contributions. \autoref{sec:results} shows the numerical results, compares conventional and PMC-based predictions, reports the $\mathrm{N^5LO}$ UHO estimates, and discusses the corresponding phenomenological implications. Finally, \autoref{sec:summary} gives the conclusions, while the LRTO results are relegated to the Appendix.
		
	\section{Theoretical Framework}
    \label{sec:th-fw}

    \subsection{pQCD corrections and the PMCs scale-setting procedure}
    
	The dominant QCD radiative correction can be written as the perturbative series
	\begin{align}\label{eq:deltaQCD}
    \delta_\text{QCD}(Q) = \sum_{i=1}^{4} r_{i}(Q,\mu_\mathrm{R}) \alpha_{s}^{i}(\mu_\mathrm{R}) + \mathcal{O}(\alpha_s^5),
    \end{align}
    where $Q$ denotes the kinematic scale of the observable, equivalently the typical momentum flow of the process, and $\mu_\mathrm{R}$ is the renormalization scale. 

    In the massless-quark approximation, the inclusive hadronic $W$-boson width is governed by the correlator of a charged non-singlet $V\!-\!A$ current. Since massless QCD is flavor-blind and chirally symmetric, the perturbative QCD corrections to this correlator coincide with those for the non-singlet vector-current correlator that appears in the standard $R$ ratio for the process $e^{+}e^{-}\to\text{hadrons}$. After the EW couplings and CKM matrix elements are factored out, the QCD correction factor $\delta_\text{QCD}$ for $W\to\text{hadrons}$ is therefore the same as the massless non-singlet $R$-ratio evaluated at $s=M_{W}^{2}$~\cite{Chetyrkin:1996ela, Baikov:2012er, dEnterria:2016rbf}. Thus, the perturbative coefficients $r_{i}(Q,\mu_\mathrm{R})$ in Eq.~\eqref{eq:deltaQCD} can be taken from Refs.~\cite{Chetyrkin:1996ela, Baikov:2008jh}. They can be parametrized through the QCD degeneracy relations among different orders~\cite{Bi:2015wea}~\footnote{We emphasize that the purpose of PMC is to solve conventional scale-setting ambiguity but not simply to improve the pQCD convergence, which is however a natural property of PMC due to elimination of divergent renormalon terms. To achieve the goal, we confirm that only the rigorous solution to the RGE can reproduce the exact $\beta$ pattern and be adopted to determine the correct magnitude of $\alpha_s$ for perturbative series~\cite{Yan:2023hra}. A recent preprint~\cite{Kotlorz:2026nkv} challenges the validity of the degeneracy relations; nevertheless, their analysis relies on an imprecise solution to the $\alpha_s$ RGE. For example, the so-called ``two-fold'' perturbative expansion they employ merely constitutes an approximate solution to the RGE~\cite{Shen:2017pdu}, and thus yields only an approximate $\beta$ pattern for the perturbative series. Consequently, both their PMC predictions and their critiques of the PMC framework are unreliable.}
    \begin{align}
     r_1 &= r_{1,0},\\
     r_2 &= r_{2,0}+\beta_0 r_{2,1},\\
     r_3 &= r_{3,0}+\beta_1 r_{2,1} + 2\beta_0 r_{3,1} + \beta_0^2 r_{3,2},\\
     r_4 &= r_{4,0}+\beta_2 r_{2,1} + 2\beta_1 r_{3,1} + 3 \beta_0 r_{4,1} \notag\\
         &\quad + \frac{5}{2}\beta_0\beta_1 r_{3,2} + 3 \beta_0^2 r_{4,2} + \beta_0^3 r_{4,3}.
    \end{align}
    Here $r_{i,0}$ are scale-invariant conformal coefficients, and $r_{i,j\neq 0}$ are scale-dependent non-conformal coefficients. The $\{\beta_{i}\}$-coefficients  of the QCD $\beta$-function have been computed up to five loops in the $\overline{\text{MS}}$-scheme~\cite{Gross:1973id, Politzer:1973fx, Gross:1973ju, Politzer:1974fr, Baikov:2016tgj, Herzog:2017ohr}. For $n_f$ active flavors, for example, $\beta_{0} = \frac{1}{4\pi}\left( \frac{11}{3}C_{A} - \frac{4}{3} T_{F} n_{f} \right)$, where $C_{A} = 3$ and $T_{F} = 1/2$ are $\text{SU}(3)_{C}$ color factors.

    Following Refs.~\cite{Brodsky:2013vpa, Shen:2017pdu, Yan:2024oyb}, the coefficients can be related to the anomalous dimension $\gamma^\mathrm{NS}$ and the polarization function $\Pi^\mathrm{NS}$~\cite{Baikov:2012zm} as~\footnote{The quantities $\gamma_i^\mathrm{NS}$ and $\Pi_i^\mathrm{NS}$ used here differ from the corresponding definitions in Ref.~\cite{Baikov:2012zm} by an overall factor $3/(4\pi^i)$. In implementing the degeneracy relations, we follow our previous studies~\cite{Brodsky:2013vpa, Shen:2017pdu, Yan:2024oyb} and retain the $n_f$-dependent terms in the anomalous dimension explicitly in the perturbative coefficients rather than absorbing them into the running coupling. This prescription differs from that adopted in Ref.~\cite{Salinas-Arizmendi:2022wrf}, where these terms are included in the scale-setting procedure.}
    \begin{align}
        r_{i,0} &= \gamma_{i}^\mathrm{NS}, \\
        r_{i,1} &= \Pi_{i-1}^\mathrm{NS}, \\
        r_{i,2} &= - \frac{\pi^{2}}{3}\gamma_{i-2}^\mathrm{NS}, \\
        r_{i,3} &= - \pi^{2} \Pi_{i-3}^\mathrm{NS}.
    \end{align}

    Following the standard PMCs scale-setting procedure~\cite{Shen:2017pdu, Yan:2022foz}, the RGE-related non-conformal $\{\beta_{i}\}$-terms are used to determine an overall effective running coupling $\alpha_{s}(Q_{*})$ for hadronic $W$-boson decay. The scale $Q_{*}$ is the PMC scale and can be interpreted as the effective momentum flow of the process. It is determined by
    \begin{align}\label{eq:PMCscale}
      \ln\frac{Q_{*}^{2}}{Q^{2}} = S_{0} + S_{1} \alpha_s(Q_{*}) + S_{2}\alpha_s^{2}(Q_{*}) + \mathcal{O}\big(\alpha_s^{3}\big),
    \end{align}
    where the explicit expressions for the coefficients $S_{i}\;(i\in\{0,1,2\})$ are given in Ref.~\cite{Yan:2024oyb}. After PMCs is implemented, the divergent renormalon contributions are removed, yielding a more convergent conformal series.
    \begin{align}\label{eq:PMCseries}
      \delta_{\mathrm{QCD}}\big|_{\mathrm{PMC}} = \sum_{i=1}^{4}r_{i,0} \alpha_s^{i}(Q_{*}) + \mathcal{O}\big(\alpha_s^{5}\big).
    \end{align}

    The PMC conformal series, being scheme- and scale-invariant~\cite{Brodsky:2012ms, Wu:2013ei, Wu:2014iba, Wu:2019mky}, provides a cleaner input for estimating the potential impact of UHO terms, thereby improving the predictive power of perturbative QCD. In the following, two probability-distribution-based methods, BA~\cite{Cacciari:2011ze, Bagnaschi:2014wea, Bonvini:2020xeo, Duhr:2021mfd, Shen:2022nyr} and LRTO~\cite{Wu:2025xdv}, will be employed to estimate the contributions from uncalculated $\mathrm{N^5LO}$ terms; a detailed comparison for the fixed-order series before and after applying the PMCs will also be presented.

    \subsection{BA estimate of UHO terms}

    The BA method provides a probability-based estimation of theoretical uncertainties from UHO terms. Detailed discussions of BA and its combination with the PMC approach can be found in Refs.~\cite{Shen:2022nyr, Shen:2023qgz, Luo:2023cpa, Yan:2023mjj, Yan:2024oyb, Yan:2022foz}. 
    
    Following the basic BA construction, for a perturbative series truncated at the $p_\mathrm{th}$ order,
    \begin{equation}\label{eq:series_eg}
        \rho_{p} = \sum_{i=1}^{p} c_{i}\alpha_{s}^{i},
    \end{equation}
    the known coefficients $\{c_1, c_2, \dots, c_p\}$ are used as prior information to constrain the next uncalculated coefficient $c_{p+1}$. The basic assumption is that, before the optimal truncation order of the asymptotic series~\footnote{The notion of the ``optimal truncation order'' was first introduced in Ref.~\cite{Beneke:1998ui}, whose value is quantified via the factorial-divergent behavior of renormalon contributions. Since the PMC eliminates the renormalon terms, the optimal truncation order -- even if still formally present -- is substantially shifted to higher orders. This significantly extends the valid predictive domain of fixed-order perturbative expansions.}, the perturbative coefficients are bounded by a common positive parameter $\bar{c}$. This parameter is inferred from the known coefficients through Bayes' theorem. After marginalizing over $\bar{c}$, one obtains a conditional probability density for $c_{p+1}$. The UHO contribution is therefore represented by a credible interval~(CI) associated with a fixed degree of belief~(DoB), rather than by a purely empirical scale-variation band. For a given DoB, the next-order prediction can be written schematically as 
    \begin{equation}\label{eq:BA_CIs}
	    \rho_{p+1}^\mathrm{(DoB)} \in \left[\rho_{p}-c_{p+1}^{(\rm DoB)}\alpha_{s}^{p+1}, \rho_{p}+c_{p+1}^{(\rm DoB)}\alpha_{s}^{p+1}\right],
    \end{equation}
    where
    \begin{align}\label{eq:BAc}
     	c_{p+1}^{(\rm DoB)}=\left\{
        \begin{aligned}
            &\bar{c}_{(p)}\frac{p+1}{p}\mathrm{DoB},\\
            &\quad\qquad\qquad\qquad\qquad\mathrm{DoB} \le \frac{p}{p+1},\\
            &\bar{c}_{(p)}\left[(p+1)(1-\mathrm{DoB})\right]^{-1/p},\\
            &\quad\qquad\qquad\qquad\qquad\mathrm{DoB} \ge \frac{p}{p+1},
        \end{aligned}\right.
    \end{align}
	with $\bar{c}_{(p)} =\max\{|c_1|, |c_2|, \dots, |c_p|\}$. For a fixed $\mathrm{DoB}$, the second branch of Eq.~\eqref{eq:BAc} applies when $p\leq \mathrm{DoB}/(1-\mathrm{DoB})$, which includes all perturbative orders considered in this work. Within this range, the factor $[(p+1)(1-\mathrm{DoB})]^{-1/p}$ decreases monotonically with $p$. Therefore, if the inclusion of a new coefficient does not increase $\bar{c}_{(p)}$, the corresponding CI progressively narrows with increasing $p$ for fixed $\alpha_s$. Conversely, a new coefficient that increases $\bar{c}_{(p)}$ may widen the subsequent CI.
    
    For convenience, we adopt $\mathrm{DoB} = 95.5\%$ as the default value to quantify the uncertainty originating from the uncomputed $\mathrm{N^5LO}$ corrections; this value matches the two-sigma ($2\sigma$) coverage probability of a two-sided normal-distribution confidence interval.

	\section{Numerical Results and Discussion}
    \label{sec:results}

    \subsection{Properties of the \texorpdfstring{$\mathrm{N}^4\mathrm{LO}$}{N4LO} QCD corrections to the hadronic \texorpdfstring{$W$}{W}-boson decay width}

    For numerical evaluation, we take the PDG inputs~\cite{ParticleDataGroup:2026aaa}: $M_W = 80.3625 \pm 0.0077\,\mathrm{GeV}$ and $\alpha_s(M_Z) = 0.1180 \pm 0.0009$. Using Eq.~\eqref{eq:deltaQCD}, the $\mathrm{N^{4}LO}$ pQCD approximant in the conventional scale-setting approach reads
    \begin{align}
     \delta_\mathrm{QCD}\big|_\mathrm{Conv.} = \left( 39.471^{+0.071+0.126}_{-0.039-0.000}\right)\times 10^{-3},
    \end{align}
    whose central value is evaluated at $\mu_{\mathrm{R}}=M_W$ with $\alpha_s(M_W)=0.1203$. Here, the strong coupling at any scale other than $M_Z$ can be obtained by using the RGE. The first uncertainty band arises from varying $\mu_\mathrm{R} \in [M_W/2,\, 2M_W]$, yielding a small net scale error of $0.28\%$; the second uncertainty accounts for the additional variation from the broader variation range $\mu_\mathrm{R} \in [M_W/4,\, 4M_W]$, which gives a larger net scale error of $0.60\%$. In principle, any value within the perturbative domain may be chosen for the renormalization scale, as this scale is unphysical. This arbitrariness gives rise to the conventional scale-setting ambiguity: truncated perturbative series exhibit strong sensitivity to the renormalization-scale choice, although such scale dependence is mitigated upon including higher-loop corrections. The error bands quoted above explicitly illustrate this ambiguity. For definiteness, we adopt the standard convention $\mu_\mathrm{R} \in [M_W/2,\, 2M_W]$ to estimate scale uncertainties in the following analysis.

    Solving Eq.~\eqref{eq:PMCscale} numerically yields the LL-, NLL-, and $\mathrm{N^2LL}$-accurate PMC scales, which correspond to the $\mathrm{N^2LO}$, $\mathrm{N^3LO}$, and $\mathrm{N^4LO}$ pQCD series, respectively:
    \begin{align}
	    Q_{*}^\mathrm{(LL,NLL,N^2LL)} = \{89.927,\, 98.929,\, 100.102\}~\mathrm{GeV}.
    \end{align}
    The results show a monotonic increase of the PMC scale: $Q_{*}^\mathrm{LL} < Q_{*}^\mathrm{NLL} < Q_{*}^\mathrm{N^2LL}$,
    with the differences between successive scale values decreasing as additional loop corrections are included. The rapid pQCD convergence of $Q_{*}$ demonstrates that the PMC scale quickly converges with increasing perturbative order. Substituting the PMC conformal series from Eq.~\eqref{eq:PMCseries} and the $\mathrm{N^2LL}$-accurate PMC scale $Q_{*}$, we obtain
    \begin{align}
	    \delta_\text{QCD}\big|_\mathrm{PMC} = 39.494 \times 10^{-3}.
    \end{align}
    This prediction is scale invariant and thus eliminates conventional scale-setting ambiguities.
    
    \begin{figure}[!t]
		\centering
		\includegraphics[width=0.48\textwidth]{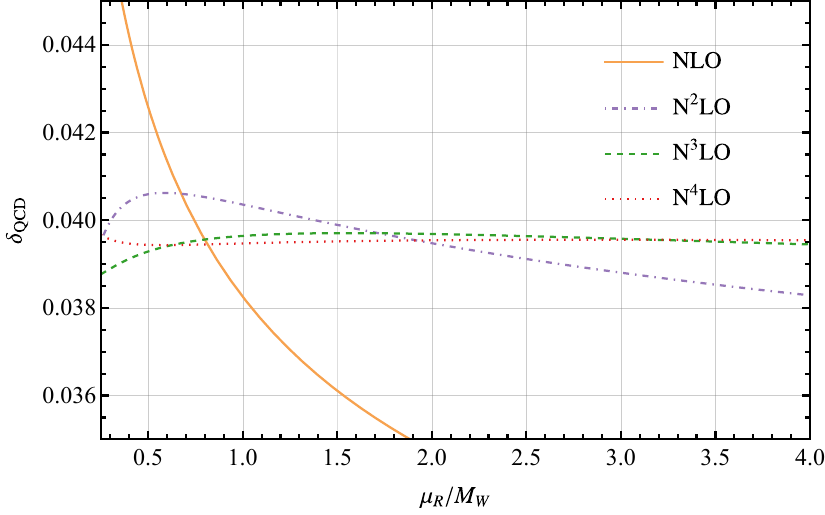}
        \includegraphics[width=0.48\textwidth]{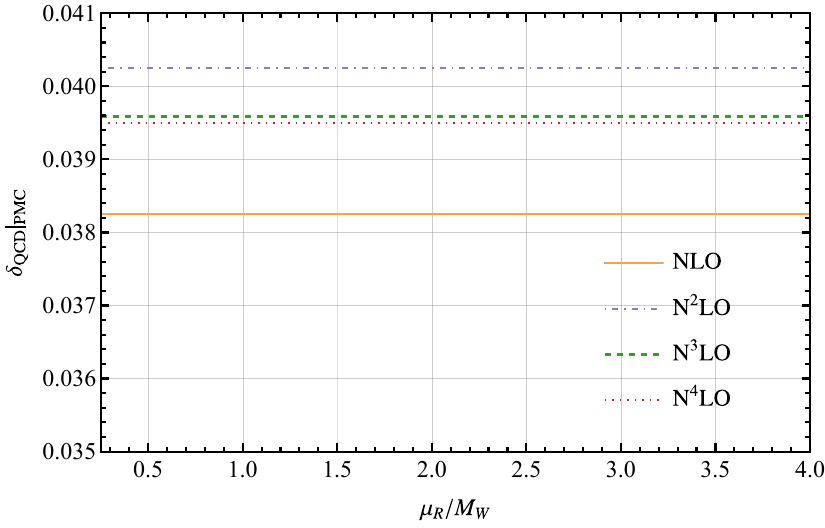}
		\caption{QCD correction $\delta_\text{QCD}$ to the hadronic $W$-boson width up to $\mathrm{N^4LO}$ as a function of the renormalization scale $\mu_\mathrm{R}$, using the conventional (upper) and PMC (lower) scale-setting approaches, respectively}
		\label{fig:Wdecay} 
	\end{figure}
   
    \begin{table*}[!t]
        \caption{Total and individual QCD corrections to the hadronic $W$-boson decay for conventional and PMC scale-setting approaches. For the conventional predictions, the central values are evaluated at $\mu_\mathrm{R} = M_W$, and the uncertainties are determined by the extrema within $\mu_\mathrm{R} \in [M_W/2,\, 2M_W]$}
        \label{tab:delta}
        \centering
        \begin{tabular}{cccccc}
            \toprule
		    & NLO & $\mathrm{N^2LO}$ & $\mathrm{N^3LO}$ & $\mathrm{N^4LO}$ & Total \\
            \midrule
		    Conv.~$(\times10^{-3})$ & $38.293 ^{+4.605}_{-3.692}$ & $2.066 ^{+2.802}_{-4.363}$ & $-0.717 ^{+0.929}_{-0.589}$ & $-0.172 ^{+0.313}_{-0.021}$ & $39.471 ^{+0.071}_{-0.039}$ \\
		    PMC~$(\times10^{-3})$ & $37.038$ & $2.525$ & $-0.049 $ & $-0.020 $ & $39.494$\\
            \bottomrule
	    \end{tabular}
	\end{table*}

    The total QCD correction $\delta_\text{QCD}$ as a function of the renormalization scale $\mu_\mathrm{R}$, before and after applying the PMC procedure, is shown in \autoref{fig:Wdecay}. The corresponding total and individual perturbative contributions up to $\mathrm{N^4LO}$ are listed in \autoref{tab:delta}. As expected, the net scale dependence of the conventional pQCD prediction diminishes as higher-order QCD corrections are included. The PMC prediction is fully independent of the initial renormalization scale and therefore reveals the intrinsic convergence pattern of the conformal series.

    \begin{figure}[htb]
		\centering
		\includegraphics[width=0.48\textwidth]{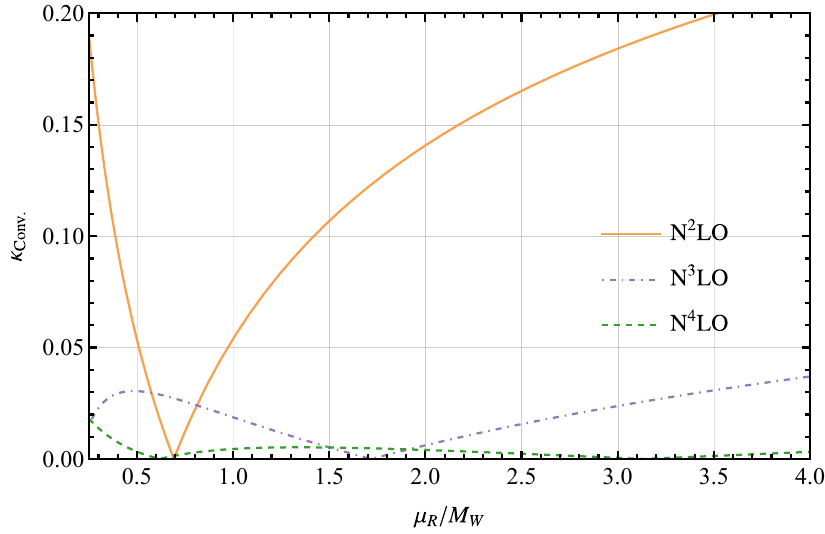}
		\caption{The $\kappa$-factor for each perturbative term ($\text{N}^2\text{LO}$, $\text{N}^3\text{LO}$, $\text{N}^4\text{LO}$) in the conventional series as a function of the renormalization scale $\mu_\mathrm{R}$}
		\label{fig:kappa} 
	\end{figure}

    The results in \autoref{tab:delta} show that the scale dependence of individual conventional perturbative terms remains sizable. The small net scale dependence of the full $\mathrm{N^4LO}$ conventional prediction arises partly from intrinsic series convergence and partly from accidental cancellations among different perturbative orders. The scale-invariant PMC series disentangles these two effects and cleanly characterizes the intrinsic perturbative behavior of the pQCD approximant. To illustrate this feature explicitly, we define the $\kappa$-factor for the $\mathrm{N^4LO}$ series from Eq.~\eqref{eq:deltaQCD} and Eq.~\eqref{eq:PMCseries} as
    \begin{align}
        \kappa_\mathrm{Conv.}^{(i)}(\mu_\mathrm{R}) &= \left|\frac{r_{i}(\mu_\mathrm{R})\,\alpha_{s}^{i} (\mu_\mathrm{R})}{r_{1}\,\alpha_{s}(\mu_\mathrm{R})}\right|, \\
        \kappa_\mathrm{PMC}^{(i)} &= \left|\frac{r_{i,0}\,\alpha_{s}^{i}(Q_{*})}{r_{1,0}\,\alpha_{s}(Q_{*})}\right|,
    \end{align}
    where $\kappa_\mathrm{Conv.}^{(i)}$ retains explicit scale dependence, as demonstrated in \autoref{fig:kappa}. The $\kappa$-factor quantifies the relative magnitude of each perturbative term and thereby characterizes the series convergence behavior. Numerically, we obtain
    \begin{align}
        \kappa_\mathrm{Conv.}(M_W/2) &=\{1,\,0.0535,\,0.0304,\,0.0033\}, \label{eq:kconv1} \\
        \kappa_\mathrm{Conv.}(M_W) &=\{1,\,0.0540,\,0.0187,\,0.0045\}, \\
        \kappa_\mathrm{Conv.}(2M_W) &=\{1,\,0.1407,\,0.0061,\,0.0040\}, \\
        \kappa_\mathrm{PMC}(\forall \mu_\mathrm{R}) &=\{1,\,0.0682,\,0.0013,\,0.0005\}.  \label{eq:kPMC} 
    \end{align}
    From Eqs.~\eqref{eq:kconv1}--\eqref{eq:kPMC}, the conventional series may exhibit good convergence for a suitable choice of $\mu_\mathrm{R}$, although its individual terms remain strongly scale-dependent. This dependence is illustrated in \autoref{fig:kappa}, which shows the $\kappa$-factor of each perturbative term as a function of $\mu_\mathrm{R}$. By contrast, the PMC series is scale-invariant and exhibits faster convergence, with the former being regarded as an intrinsic characteristic expected of a physical prediction.
		
	\subsection{Estimation of N\texorpdfstring{$^5$}{5}LO QCD corrections using the BA approach}
    
    In this subsection, we present the fixed-order values of $\delta_\mathrm{QCD}$ up to $\mathrm{N^4LO}$, alongside the CIs predicted by the probability-based BA approach, for two representative DoB choices, $95.5\%$ and $99.7\%$.
    
    \begin{figure}[htb]
		\centering
		\includegraphics[width=0.48\textwidth]{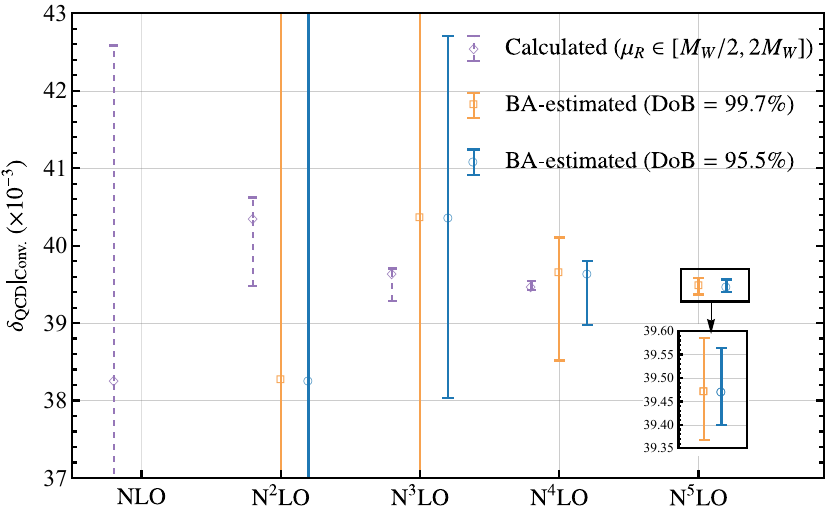}
        \includegraphics[width=0.48\textwidth]{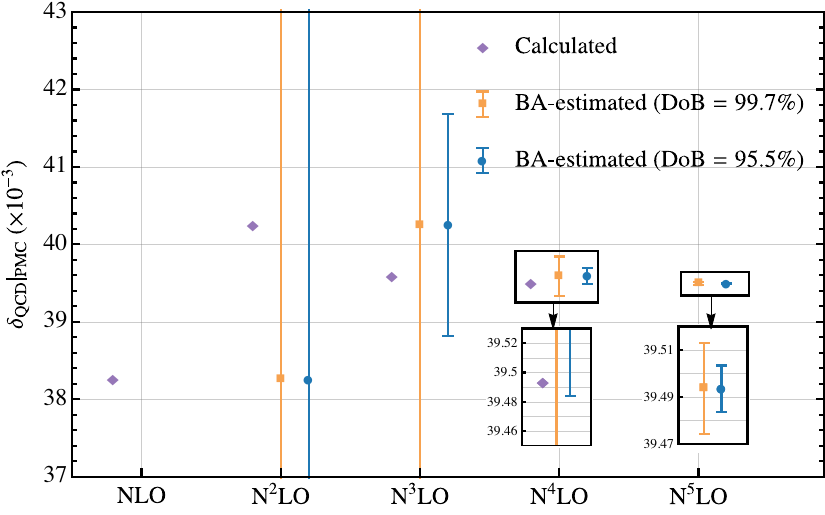}
		\caption{Comparison of the BA-predicted CIs, obtained using fixed-order theoretical values up to $\mathrm{N^5LO}$ under conventional (upper) and PMC (lower) scale-setting approaches for two typical choices of the DoB. Hollow and solid diamonds denote the calculated fixed-order pQCD values obtained with conventional and PMC scale-setting approaches, respectively. Squares and circles with error bars denote BA-predicted CIs based on the known conventional and PMC series, respectively, for different DoB values}
		\label{fig:UHOBA}
	\end{figure}
    
    As shown in \autoref{fig:UHOBA}, the BA-estimated CIs generally shrink as additional perturbative orders are incorporated, reflecting the iterative Bayesian update of the underlying coefficient sequence. For the conventional series, the CIs are constructed by taking the envelope over $\mu_\mathrm{R}\in[M_W/2,\,2M_W]$, and the scale dependence renders them asymmetric. At a given DoB, the CIs derived from the scale-invariant PMC series are narrower than their conventional counterparts, particularly at higher orders. This observation highlights the importance of using the more precise scale-independent conformal series as the basis for estimating the UHO contribution.
    
    For the fully known $\mathrm{N^4LO}$ expansion, the BA estimates of the UHO uncertainties associated with the $\mathrm{N^5LO}$ contribution at $\mathrm{DoB} = 95.5\%$ are
    \begin{align}
      \Delta\delta_\text{QCD}\big|_\mathrm{Conv.}^\mathrm{BA} &= \binom{+9.3}{-7.1} \times 10^{-5},\\
      \Delta\delta_\text{QCD}\big|_\mathrm{PMC}^\mathrm{BA} &= \pm 1.0 \times 10^{-5}.
    \end{align}
    At $\mathrm{DoB}=99.7\%$, the corresponding estimates are
    \begin{align}
       \Delta\delta_\text{QCD}\big|_\mathrm{Conv.}^\mathrm{BA} &= \binom{+11.4}{-10.2} \times 10^{-5},\\
       \Delta\delta_\text{QCD}\big|_\mathrm{PMC}^\mathrm{BA} &= \pm 1.9 \times 10^{-5}.
    \end{align}
    \autoref{fig:UHOBA} shows that both the conventional and PMC series pass all available order-by-order coverage tests at $\mathrm{DoB}=95.5\%$ and $99.7\%$. At each testable order, the calculated fixed-order central value lies within the corresponding BA CI inferred from information available only through the preceding order. This consistent coverage supports the reliability of the BA extrapolation to the uncalculated $\mathrm{N^{5}LO}$ contribution for the present observable.

    \subsection{Phenomenological implications}

    The hadronic decay width of the $W$ boson can be calculated via Eq.~\eqref{eq:Gamma}. For the CKM matrix elements $V_{qq'}$, we use the parametrization of Ref.~\cite{Harari:1986xf}, as advocated by the PDG, and take the following parameter values consistent with PDG~\cite{ParticleDataGroup:2026aaa}:
    \begin{align}
        \sin\theta_{12} &= 0.22517, &\sin\theta_{13}& = 0.003763, \notag\\
        \sin\theta_{23} &= 0.04189, &\delta &= 1.154,
    \end{align}
    which enforce the unitarity of the CKM matrix, namely, $\sum\limits_{q=u,c} \sum\limits_{q'=d,s,b} |V_{qq'}|^2 = 2$.

    \begin{table}[htb]
		\caption{\label{tab:err}
        Additional uncertainties (in units of MeV) induced by $\Delta\alpha_s(M_Z)=\pm 0.0009$, $\Delta M_W=\pm 0.0077~\mathrm{GeV}$, and the BA estimate of the UHO contribution at $\mathrm{DoB}=95.5\%$}
        \centering
        \begin{tabular}{cccc}
            \toprule
            & $\Delta\Gamma_\text{QCD}|^{\Delta\alpha_{s}}$
            & $\Delta\Gamma_\text{QCD}|^{\Delta M_W}$ 
            & $\Delta\Gamma_\text{QCD}|^\mathrm{UHO}$ \\
            \midrule
			Conv. & $\pm 0.4199$ & $^{+0.4064}_{-0.4063}$ & $^{+0.1267}_{-0.0963}$ \\
			PMC   & $\pm 0.4212$ & $^{+0.4064}_{-0.4063}$ & $\pm 0.0134$\\
            \bottomrule
		\end{tabular}
	\end{table}
    
    When only QCD corrections are included, we obtain the corresponding $\mathrm{N^4LO}$ predictions:
    \begin{align}
        \Gamma_{W,\text{QCD}}^\mathrm{had}|_\mathrm{Conv.} &= 1416.27 \pm 0.60 ~\mathrm{MeV},\\
        \Gamma_{W,\text{QCD}}^\mathrm{had}|_\mathrm{PMC} &= 1416.30 \pm 0.59 ~\mathrm{MeV},
    \end{align}
    where the uncertainties are combined in quadrature with those originating from $\Delta\alpha_s(M_Z)=\pm 0.0009$, $\Delta M_W=\pm 0.0077\,\mathrm{GeV}$, and from the UHO contributions estimated via the BA approach at $\mathrm{DoB}=95.5\%$. Each uncertainty is shown in \autoref{tab:err}. \autoref{tab:err} shows that the uncertainties induced by $\Delta\alpha_s(M_Z)$ and $\Delta M_W$ are nearly identical for both the conventional and PMC predictions, since the corresponding central values are very close. In contrast, the UHO uncertainty exhibits a distinct behavior. In the conventional perturbative series, the BA estimate yields an uncertainty of $\mathcal{O}(0.1\,\mathrm{MeV})$, comparable to the residual uncertainty from scale variation at fixed order. For the PMC series, the corresponding BA uncertainty is reduced to $\mathcal{O}(0.01\,\mathrm{MeV})$. Since the UHO contributions are considerably smaller than the other error sources, the choice of $\mathrm{DoB}$ only slightly affects the total uncertainties; for instance, increasing the $\mathrm{DoB}$ to $99.7\%$ changes the PMC UHO component to $\pm 0.0264\,\mathrm{MeV}$, while the total uncertainty remains approximately $\pm 0.59\,\mathrm{MeV}$. Accordingly, we adopt $\mathrm{DoB}=95.5\%$ for all subsequent discussions.
    
    We further update the pure EW and mixed QCD--EW corrections by including finite-fermion-mass effects using the formulae of Refs.~\cite{Denner:1990cpz, Denner:1991kt, Bohm:1986rj}. This gives $\Delta\Gamma_\mathrm{EW}=-4.96\,\mathrm{MeV}$ and $\Delta\Gamma_\mathrm{mix}=-0.76\,\mathrm{MeV}$ within the $G_F$-scheme. Including also the quark-mass contribution to the QCD correction, $-1.65\,\mathrm{MeV}$, we obtain
    \begin{align}
        \Gamma_{W,\text{Total}}^\mathrm{had}|_\mathrm{Conv.} &= 1408.91 \pm 0.60\,\mathrm{MeV},\\
        \Gamma_{W,\text{Total}}^\mathrm{had}|_\mathrm{PMC} &= 1408.94 \pm 0.59\,\mathrm{MeV}.
    \end{align}
        
\begin{figure}[htb]
        \centering
        \includegraphics[width=0.48\textwidth]{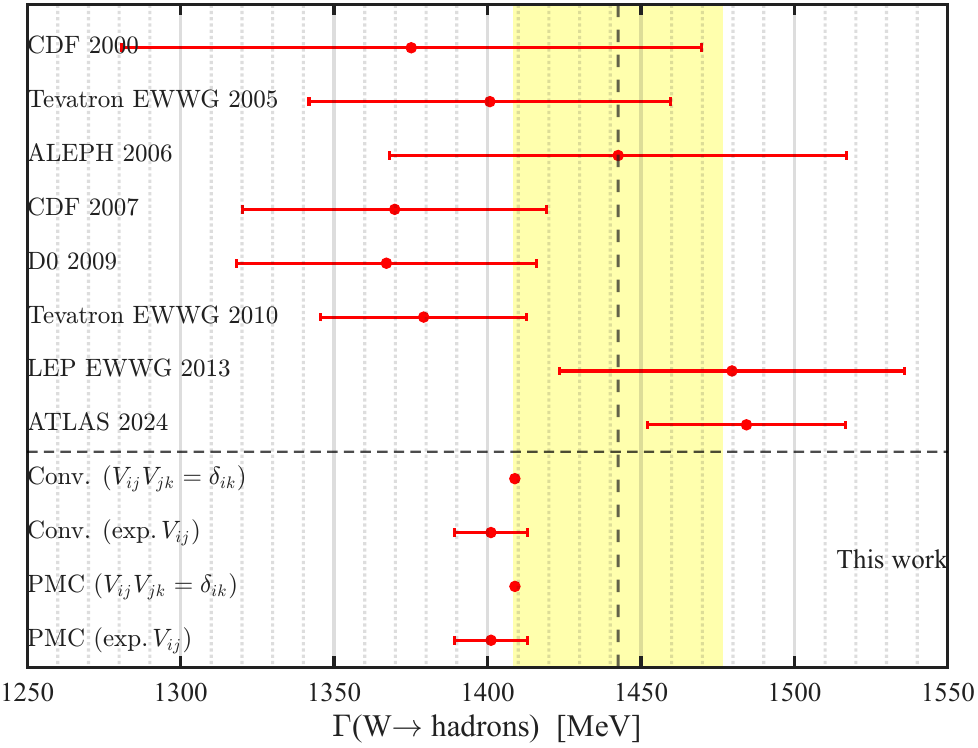}
		\caption{Comparison of the product of the $W$-boson total decay width and hadronic branching fraction from recent experimental measurements (red points with error bars), the PDG average value (dashed line), and the theoretical result obtained in the present work. The experimentally determined CKM matrix-element combination quoted by the PDG is $\sum\limits_{q=u,c}\sum\limits_{q'=d,s,b}|V_{qq'}|^2=1.988 \pm 0.017$~\cite{ParticleDataGroup:2026aaa}}
		\label{fig:comp_exp}
	\end{figure}

    The comparison in \autoref{fig:comp_exp} shows recent experimental measurements of the hadronic $W$-boson decay width \cite{CDF:2000hii, Ashmanskas:2005ij, ALEPH:2006cdc, CDF:2007tdb, D0:2009oet, TevatronElectroweakWorkingGroup:2010mao, ALEPH:2013dgf, ATLAS:2024erm} with the result obtained in this work. Under CKM unitarity, the prediction agrees with the measurements within their current uncertainties. If unitarity is not imposed, the normalization depends directly on $\sum_{qq'}|V_{qq'}|^2$, whose uncertainty must then be propagated.
    
    The hadronic branching ratio $\mathcal{B}$ of the $W$ boson is defined as the ratio of the hadronic decay width $\Gamma_{W}^\mathrm{had}$ to the total decay width $\Gamma_{W}$:
	\begin{equation}
		\mathcal{B}(W\to \text{hadrons}) = \frac{\Gamma_{W}^\mathrm{had}}{\Gamma_{W}}.
	\end{equation}

    Using the input values $\Gamma_{W}=2140\pm 50~\mathrm{MeV}$~\cite{ParticleDataGroup:2026aaa} and $ \Gamma_{W}^\mathrm{had}|_\mathrm{PMC} =1408.94\pm 0.59~\mathrm{MeV}$, the final hadronic branching ratio is~\footnote{The errors are calculated through the usual error propagation formula, that is, the error of a quantity $Z = X/Y$ is $\Delta Z=(X_{0}/Y_{0})\sqrt{(\Delta X/X_{0})^{2}+(\Delta Y/Y_{0})^{2}}$, where $X=X_0 \pm \Delta X$, $Y=Y_0 \pm \Delta Y$, and $Z=(X_0/Y_0) \pm \Delta Z$. }
	\begin{equation}
		\mathcal{B}(W\to \text{hadrons})|_\mathrm{PMC} = (65.84\pm 1.54)\%.
	\end{equation}

    \section{Summary}
    \label{sec:summary}

    In this work, we investigated the massless QCD corrections to the hadronic decay width of the $W$ boson up to $\mathrm{N^4LO}$, and estimated the uncalculated $\mathrm{N^5LO}$ contribution. Although the conventional $\mathrm{N^4LO}$ prediction already exhibits good perturbative convergence and small net scale dependence over $\mu_\mathrm{R} \in [M_W/2,\, 2M_W]$, its individual perturbative terms remain highly sensitive to the renormalization scale. This order-by-order scale sensitivity introduces an additional source of ambiguity into probability-based estimates of UHO contributions. We therefore applied the PMCs to obtain a scale-independent conformal series with improved convergence. Using the known $\mathrm{N^4LO}$ series, we obtained $Q_{*}^{\mathrm{N^2LL}} = 100.102~\mathrm{GeV}$ and $\delta_\mathrm{QCD}|_\mathrm{PMC} = 39.494\times 10^{-3}$. We then employed BA approach to estimate the $\mathrm{N^5LO}$ contribution, which yields a $95.5\%$ CI of $\pm 1.0\times 10^{-5}$.

    With the parametric uncertainties associated with $\alpha_s(M_Z)$ and $M_W$ included, the QCD-corrected hadronic width is $\Gamma_{W,\mathrm{QCD}}^\mathrm{had}|\mathrm{PMC}=1416.30\pm0.59~\mathrm{MeV}$. After further incorporating the pure EW, mixed QCD--EW, and quark-mass corrections, we obtain the total hadronic decay width, $\Gamma_{W,\mathrm{Total}}^\mathrm{had}|_\mathrm{PMC}=1408.94\pm0.59~\mathrm{MeV}$. The BA uncertainty associated with the uncalculated $\mathrm{N^5LO}$ term amounts to $0.0134~\mathrm{MeV}$ and is therefore subdominant to the combined parametric uncertainty. The primary benefit of PMC in this analysis is the construction of a scale-invariant and more rapidly convergent perturbative series, which provides a consistent basis for probability-based estimates of UHO contributions. Such a series establishes a robust perturbative foundation for future high-precision studies of $W$-boson decay observables.

    \backmatter
    
    \bmhead{Acknowledgements}
    The authors would like to thank Xu-Chang Zheng for helpful discussions.
    \section*{Declarations}
    
    \begin{itemize}
        \item \textbf{Competing interests} The authors declare that they have no relevant financial or non-financial interests to disclose.
        \item \textbf{Data availability}  Data sharing is not applicable to this article as no datasets were generated or analysed during the current study. All numerical results can be obtained from the formulas and input parameters given in the manuscript.
        \item \textbf{Funding} This work was supported in part by the Natural Science Foundation of China under Grant Nos. 12547115, 12575080 and 12547101.
        \item \textbf{Code availability} Code/Software sharing is not applicable to this article as no code/software was generated or analysed during the current study.
        \item \textbf{Author contributions} All authors contributed to the conception, analysis, interpretation of the results, and preparation of the manuscript. All authors read and approved the final manuscript.
    \end{itemize}
    
    \setcounter{mainfigurecounter}{\value{figure}}

    \begin{appendices}
    \setcounter{figure}{\value{mainfigurecounter}}
    \renewcommand{\thefigure}{\arabic{figure}}
    
    \section{Results and discussion of the LRTO}
    In this appendix, we present the results for another probability-distribution-based statistical approach,  linear regression through the origin (LRTO).

    \subsection{Formulas for the LRTO estimate of UHO terms}
    \label{sec:LRTO_th}

    As an independent estimate of the UHO contribution, we also employ the LRTO method, recently proposed in Ref.~\cite{Wu:2025xdv}. This method exploits the asymptotic behavior of pQCD expansions prior to the optimal truncation order. In this regime, the magnitude of each perturbative term is predominantly governed by the power suppression of $\alpha_s$, while subleading deviations from this scaling trend are treated as a source of theoretical uncertainty. The method introduces a $K$-factor for the normalized series defined in Eq.\eqref{eq:series_eg}~\footnote{For the LRTO method, a full pQCD series takes the form $\tilde{\rho}_{p} = c_{0}\alpha_{s}^{l}\bigl(1+\rho_{p}\bigr)$ with integer $l\ge 0$.}:
    \begin{equation}\label{eq:K_LRTO}
        K_{k} = c_{k}\alpha_{s}^{k}.
    \end{equation}
    One then assumes that the magnitude of $|K_k|$ follows an approximately exponential behavior,
    \begin{equation}
        |K_k|=u^k \exp(\epsilon_k),
    \end{equation}
    where the parameter $u$ characterizes the convergence rate and $\epsilon_k$ denotes residual subleading fluctuations. Taking the logarithm gives
    \begin{align}
		\ln |K_k| &= k\,\theta + \epsilon_k, & \theta&=\ln u.
	\end{align}
    Thus, estimating the trend of the UHO terms is reduced to fitting the slope $\theta$ of a straight line through the origin. The best-fit value $\hat{\theta}_{p}$ is
    \begin{align}
        \hat{\theta}_{p} = \frac{6}{(2p+1)(p+1)p}\sum_{k=1}^{p}k \ln|K_k|,
    \end{align}
    which can be extrapolated to predict the magnitude of $K_{p+1}$ and hence the next UHO contribution:
    \begin{align}
		\rho_{p+1}^{(\mathrm{DoB})} &\in \left[\rho_{p}-\Delta_{p+1}^\mathrm{(DoB)},\rho_{p}+\Delta_{p+1}^\mathrm{(DoB)}\right]
	\end{align}
    with
    \begin{align}
        \Delta_{p+1}^\mathrm{(DoB)} = \exp\left[(p+1)\left(\hat{\theta}_{p} + z_\mathrm{DoB}\delta_{p}\right) \right],
    \end{align}
    where $z_\mathrm{DoB} = 1, 2$, and $3$ for $\mathrm{DoB} = 68.3\%, 95.5\%$, and $99.7\%$, respectively; and
    \begin{align}
        \delta_{p} &= \sqrt{\frac{6}{(2p+1)(p+1)p} \hat{\sigma}_{p}^{2}},\\
        \label{LRTO:sigma}
        \hat{\sigma}_{p}^{2} &= \frac{1}{p-1}\sum_{k=1}^{p} \left(\ln|K_k|-k\,\hat{\theta}_{p}\right)^2.
    \end{align}
    
    The LRTO method therefore provides a direct measure of the convergence behavior of the perturbative series. Its reliability can be assessed via the fitted uncertainty and goodness of fit. Like the BA method, it is expected to yield more stable behavior when applied to the scale-invariant, better-convergent PMC series. Similarly, as our default choice, we set $\mathrm{DoB} = 95.5\%$, or equivalently $z_\mathrm{DoB} = 2$, to estimate the UHO uncertainty.

    It is worth mentioning that, within conventional perturbative series, the coefficients $r_{i>1}$ depend explicitly on the chosen renormalization scale. The BA and LRTO methods must therefore be applied only after fixing a renormalization scale, which introduces an additional convention into the UHO uncertainty estimate, as will be demonstrated in Figs.~\ref{fig:UHOBA} and \ref{fig:UHOLRTO}. By contrast, the conformal coefficients $r_{i,0}$ appearing in the PMC series are scale independent and provide a cleaner coefficient sequence as input for the UHO uncertainty evaluation.
 
    \subsection{Results of LRTO}
    
    \begin{figure}[htb]
		\centering
		\includegraphics[width=0.48\textwidth]{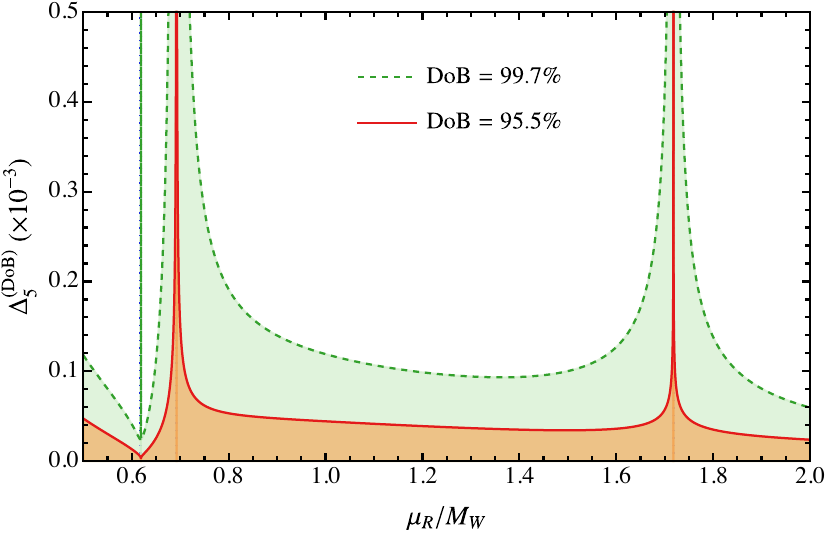}
		\caption{LRTO estimates of the $\mathrm{N^5LO}$ contribution as a function of the renormalization scale under conventional scale-setting approach. The green and red bands correspond to $\mathrm{DoB}=99.7\%$ and $95.5\%$, respectively}
		\label{fig:LRTO_div} 
	\end{figure}
    
    \begin{figure*}[htb]
		\centering
		\includegraphics[width=0.48\textwidth]{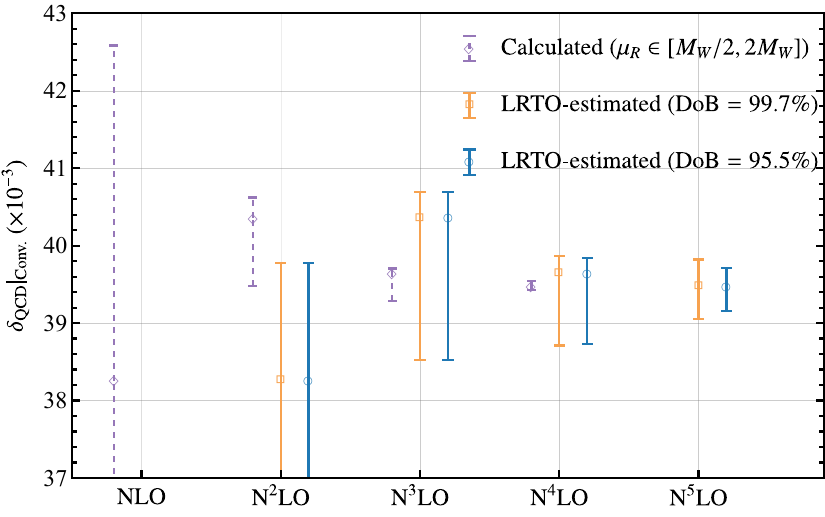}
        \includegraphics[width=0.48\textwidth]{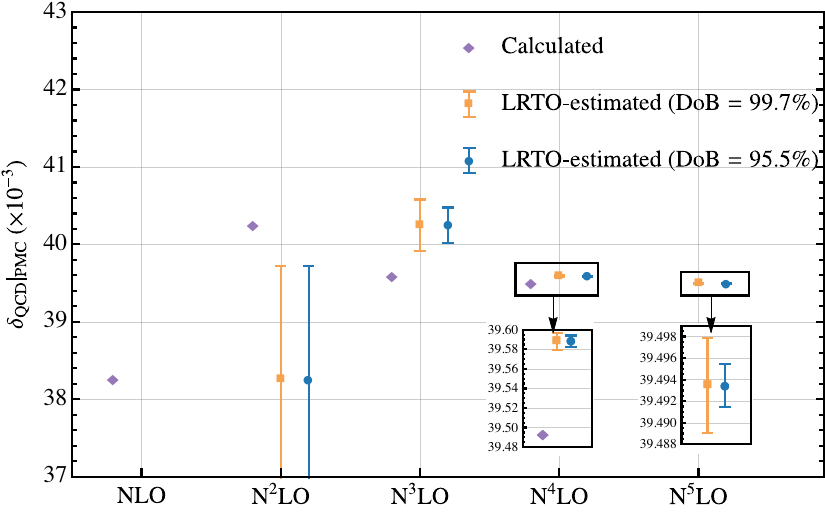}
		\caption{Comparison of the LRTO-predicted CIs, obtained using fixed-order theoretical values up to $\mathrm{N^5LO}$ under conventional (Conv., left) and PMC (right) scale-setting approaches for two typical choices of the DoB. Hollow and solid diamonds denote the calculated fixed-order pQCD values obtained with conventional and PMC scale-setting approaches, respectively. Squares and circles with error bars denote LRTO-predicted CIs based on the known conventional and PMC series, respectively, for different DoB values}
		\label{fig:UHOLRTO}
	\end{figure*}

    We first give the results for the conventional series. As shown in \autoref{fig:kappa}, scale variation can drive an individual contribution \(K_k(\mu_\mathrm{R})\) to zero at particular values of \(\mu_\mathrm{R}\). Because LRTO is formulated in terms of \(\ln|K_k|\), each zero produces the logarithmic singularity \(\ln|K_k|\to-\infty\) and destabilizes the inferred UHO contribution. Consequently, the direct procedure described in \autoref{sec:LRTO_th} cannot provide a finite, stable estimate over the full scale interval, as illustrated in \autoref{fig:LRTO_div}.

    To nevertheless provide an LRTO-based estimate while retaining the scale dependence of the conventional series, we employ the coefficient of determination discussed in Ref.~\cite{Wu:2025xdv},
    \begin{align}
        \mathcal{R}^2 =
        \frac{\sum_{k=1}^{p} k^2\hat{\theta}^{2}}
        {\sum_{k=1}^{p}\ln^2|K_k|}.
    \end{align}
    This statistic quantifies how closely the known terms follow the approximate linear relation between $\ln|K_k|$ and $k$ assumed by LRTO. Because the scale that maximizes $\mathcal{R}^2$ is generally not unique, we first identify all such scales within the prescribed interval. LRTO is then applied to each associated series, and every estimate is evolved across the scale interval using the RGE. The outer envelope of these evolved estimates defines the LRTO CI for the conventional prediction. This construction is a conservative workaround rather than a fully satisfactory direct application of LRTO. As shown in the left panel of \autoref{fig:UHOLRTO}, the combination of scale selection, RGE evolution, and envelope construction substantially broadens the resulting CIs. These intervals provide little meaningful constraint on the first omitted contribution and are therefore unsuitable for precision applications. This outcome is consistent with Ref.~\cite{Wu:2025xdv}, where a conventional-series LRTO estimate was not pursued.

    Applying LRTO to the PMC series does not encounter these scale-induced singularities but reveals a different limitation. As shown in the right panel of \autoref{fig:UHOLRTO}, the LRTO-estimated CIs fail to cover the corresponding calculated PMC values in every available order-by-order backtest. For the uncalculated $\mathrm{N^5LO}$ contribution, the estimates at the nominal level $\mathrm{DoB}=95.5\%$ \footnote{Under the LRTO convention of Ref.~\cite{Wu:2025xdv}, the actual probability content of an interval labeled by the nominal DoB is $\mathrm{DoB}_{\mathrm{act}}=50\%+\mathrm{DoB}/2$. Matching the actual BA levels of $95.5\%$ and $99.7\%$ would therefore require nominal LRTO inputs of $91.0\%$ and $99.4\%$, respectively, producing narrower intervals. Because the current LRTO intervals already fail all available PMC backtests, this recalibration would only strengthen the observed coverage failure. We retain the nominal DoB convention of Ref.~\cite{Wu:2025xdv} for consistency.} are
    \begin{align}
       \Delta\delta_\text{QCD}\big|_\mathrm{Conv.}^\mathrm{LRTO} &= \binom{+24.6}{-31.4} \times 10^{-5},\\
       \Delta\delta_\text{QCD}\big|_\mathrm{PMC}^\mathrm{LRTO} &= \pm 0.2 \times 10^{-5}.
    \end{align}
    At $\mathrm{DoB}=99.7\%$, the corresponding estimates are
    \begin{align}
       \Delta\delta_\text{QCD}\big|_\mathrm{Conv.}^\mathrm{LRTO} &= \binom{+35.1}{-41.2} \times 10^{-5},\\
        \Delta\delta_\text{QCD}\big|_\mathrm{PMC}^\mathrm{LRTO} &= \pm 0.4 \times 10^{-5}.
    \end{align}
    Taken together, \autoref{fig:UHOBA} and \autoref{fig:UHOLRTO} show that the narrow PMC LRTO CIs should not be interpreted as evidence of greater precision. The order-by-order failures indicate that the known terms of the PMC series do not support LRTO’s assumption that $\ln|K_k|$ depends approximately linearly on $k$. Such log-linearity in turn requires an approximately stable geometric convergence rate, not merely convergence of the perturbative series. LRTO is therefore not sufficiently reliable to provide the nominal UHO uncertainty for this observable. 
    
    For the PMC series, LRTO yields a $95.5\%$ CI of $\pm 0.2\times 10^{-5}$, which is narrower than BA one. However, LRTO fails all available order-by-order coverage tests for the PMC series and becomes ill-defined for the conventional series at renormalization scales where individual perturbative contributions vanish. The coverage failures show that the present observable does not support the approximately log-linear convergence pattern assumed by LRTO, whereas the singularities reflect a structural limitation of its dependence on $\ln|K_k|$. We therefore adopt the BA result to do our analysis of the branching ratios in the body of the text. 
    
    \end{appendices}
    
    \bibliography{ref}

\end{document}